\documentclass[a4paper,10pt]{article}

\usepackage{mathrsfs,latexsym,amsmath,amssymb,amsthm}
\usepackage{setspace}
\usepackage{cite}

\usepackage{color,cite,graphicx}

\newtheorem{theorem}{Theorem}
\newtheorem{rmk}{Remark}

\input xypic
\title{Dynamics of Solutions of $f(R)$ Theory}
\date{}
\author{Lucy MacNay\\Max-Planck Institute for Gravitational Physics\\Am M\"{u}hlenberg 1\\Golm, 14476\\ Germany}

\begin{document}

\maketitle

\begin{abstract}
 We use the correspondence between the $f(R)$ theory and an Einstein-scalar field system to study late-time dynamics of solutions of $f(R)$ theory.  We discuss how reasonable assumptions on the potential of the scalar field lead to restrictions on the function $f(R)$ and use known results for the scalar field system to gain results on solutions of the $f(R)$ theory.  In particular, we prove accelerated expansion at late times for several different categories of functions $f(R)$ that satisfy certain restrictions.
\end{abstract}

\section{Introduction}

Several years ago it was discovered, from the studies of the supernovae data, that the Universe is undergoing accelerated expansion and since this time, there has been a renewed interest in modified gravity models.  There is now such a vast amount of evidence for the acceleration of the Universe (supernovae data \cite{supernovae}, large scale structure formation \cite{lss}, cosmic microwave background radiation \cite{cmb}, weak lensing \cite{wl}, etc.) that this has led to a great deal of research into explanations for the driving force behind this accelerated expansion.  Some of the simpler theories involve introducing a cosmological constant or some kind of dark energy, perhaps driven by a scalar field.  However, it may be possible to explain the accelerated expansion by modifying Einstein's theory of relativity itself.  These theories are known as ``modified gravity'' or ``higher-order gravity'' theories and the particular case we are concerned with here is that of ``$f(R)$ gravity'', where we replace the scalar curvature, $R$, in the Lagrangian of Einstein's theory of relativity with an arbitrary function, $f(R)$, of $R$.  

The idea of using a non-linear $R$ is by no means a modern one.  It became of interest in the study of quantum gravity in the 60's and has greatly increased in popularity in the last 10 years.  There have been many recent papers discussing the viability of $f(R)$ theories with respect to solar system tests and cosmological constraints (for a sample, see \cite{carroll, faulkner,starobinsky, HuSawicki,tsujikawa, Appleby, fay, amendola, nojiri} and references therein). 

Although the literature on $f(R)$ theories is vast, the portion of papers that study $f(R)$ theory from a \emph{mathematical} viewpoint is much smaller and, although it is believed that $f(R)$ theories may explain the observed accelerated expansion of the universe, there are few rigorous mathematical proofs that this is indeed the case.  This paper will provide some such rigorous proofs for certain cases of $f(R)$ theory.  We will not consider a particular form of the function $f(R)$, but rather prove results on the late-time dynamics of $f(R)$ theories under certain restrictions on the function $f(R)$.

Since the field equations of the $f(R)$ theory are 4th order partial differential equations, this makes the theory analytically complicated.  However, the theory can be related, via a conformal transformation, to Einstein's theory of relativity where the matter is governed by a scalar field.  Following a suggestion given in \cite{late-time}, we can exploit this relationship to gain results on the dynamics of solutions to the $f(R)$ theory using known results on the dynamics of solutions to the Einstein-scalar field system.  

The layout of the paper is as follows: In section 2, we give an introduction to and derive the field equations for the $f(R)$ theory.  Section 3 will deal with the vacuum $f(R)$ case.  We will first recap the relationship between the $f(R)$ theory and the Einstein-scalar field system.  We will then consider two different restrictions on the potential of the scalar field, for which the late-time dynamics of the solutions are known \cite{Rendallposmin, slowroll}.  We will translate these restrictions into restrictions on the function $f(R)$ and prove results on the late-time dynamics of the $f(R)$ theory.  In section 4, we will consider $f(R)$ theory with ordinary matter satisfying the strong and dominant energy conditions.  We will first describe a relationship between this $f(R)$-matter system and the coupled Einstein-scalar field-matter system studied in \cite{Bieli}.  We will then use results in \cite{Bieli} on the late-time dynamics of the Einstein-scalar field-matter system where the potential is restricted in the same way as in the vacuum case to prove results on the late-time dynamics of the $f(R)$-matter system. 

We will be dealing with cosmological models that are homogeneous and isotropic.  The underlying spacetimes are the Friedmann-Lema\^{\i}tre-Robertson-Walker (FLRW) models.  If we assume that the metric of the slices of constant time is flat, then the metric takes the form $$ ds^2 = -dt^2 + a^2(t)(dx^2 + dy^2 + dz^2) $$ where $a(t)$ is the scale factor.  The scale factor is important for proving accelerated expansion, since it gives a measurement of distances between galaxies.  Therefore $\dot{a} > 0$ means that the universe is expanding, while $\ddot{a} > 0$ gives accelerated expansion of the universe.  Another useful parameter is the Hubble parameter, $H$, which is defined by $$ H = \frac{\dot{a}}{a} $$  We assume $c = G = 1$. 

\section{$f(R)$ theory}

It is well known that the Einstein field equations which describe Einstein's Theory of Relativity can be derived from variational principles (see, for example \cite{Wald}).  To do this, one considers the Einstein-Hilbert action
\begin{equation}
 \mathcal{L} = \int R \sqrt{-g}d^4x
\end{equation}
where $R$ is the scalar curvature and $g$ is the determinant of the metric.  Taking the variation of this action with respect to the metric, $g_{\alpha\beta}$, yields the vacuum Einstein equations
\begin{equation}
 R_{\alpha\beta} - \frac{1}{2}Rg_{\alpha\beta} = 0
\end{equation}
where $R_{\alpha\beta}$ is the Ricci tensor.  

We can also add matter to the Einstein equations by choosing a Lagrangian of the form
\begin{equation} \label{matteraction}
 \mathcal{L} = \int \left[R + 8\pi L^M \right] \sqrt{-g}d^4x
\end{equation}
where $L^M$ is the Lagrangian density for the matter field.  Then the Einstein equations are given by
\begin{equation}
 R_{\alpha\beta} - \frac{1}{2}Rg_{\alpha\beta} = 8\pi T^M_{\alpha\beta}
\end{equation}
where the energy-momentum tensor, $T^M_{\alpha\beta}$ is defined by
\begin{equation}
 T^M_{\alpha\beta} = -\frac{\partial L^M}{\partial g^{\alpha\beta}} + \frac{1}{2}L^Mg_{\alpha\beta}
\end{equation}

The matter field must also satisfy equations of motion, which are found by taking the variational derivative of the action (\ref{matteraction}) with respect to the matter field.  In their general form they are given by
\begin{equation}
 \frac{\partial L^M}{\partial \phi} - \nabla^\alpha \left(\frac{\partial L^M}{\partial (\nabla^\alpha \phi)}\right) = 0
\end{equation}
where $\phi$ is the matter field.  There are many possible choices for the matter and therefore for the energy-momentum tensor.  Some describe real, physical matter, while others are mathematical tools.  For more information on matter models see, for example, \cite{Rendallbook}.

As previously stated, experimental results over the last ten years have shown that the universe is undergoing accelerated expansion.  There are several ways of adapting Einstein's theory of relativity to account for this accelerated expansion.  The simplest way is to introduce a cosmological constant to the left-hand-side of the Einstein equations.  Another, more sophisticated way is to include matter in the form of a scalar field.  

However, in this paper, we are interested in an alternative way of explaining this accelerated expansion.  We will consider the idea of modifying gravity to build more freedom into the theory itself.  In the study of $f(R)$ theories, we replace the $R$ in the Einstein-Hilbert action with a general function of $R$, to get the action
\begin{equation}
 \mathcal{L} = \int  f(R) \sqrt{-g} d^4x
\end{equation}
This then gives rise to the \emph{vacuum $f(R)$ field equations}
\begin{equation} \label{FE}
 f'(R) R_{\alpha\beta} - \frac{1}{2}f(R)g_{\alpha\beta} - \nabla_\alpha \nabla_\beta (f'(R)) + \square (f'(R)) g_{\alpha\beta} = 0
\end{equation}
where a dash denotes differentiation with respect to $R$, $\nabla_\alpha$ is covariant differentiation with respect to the metric $g_{\alpha\beta}$ and $\square = \nabla_\alpha \nabla^\alpha$ is the d'Alembertian operator.

\section{The $f(R)$ theory in vacuum}

\subsection{$f(R)$ and the Einstein-scalar field system}

The $f(R)$ field equations are 4th order in the metric, compared to the 2nd order Einstein equations, which makes them harder to study.  Fortunately, we can use a conformal transformation to transform them into a system of equations which are 2nd order in the metric and then gain results from this simpler set of equations, which turn out to be Einstein equations with a specific type of matter.  

A reference for this work is given in \cite{Mukhanov}, although it should be noted that the signature used in the reference is $(+---)$, compared to the $(-+++)$ signature we use here.

The conformal transformation that we use is 
\begin{equation} \label{metrics}
g \mapsto \tilde{g} \hspace{.2in} \textrm{ where } \hspace{0.2in} \tilde{g}_{\alpha\beta} = Fg_{\alpha\beta} \end{equation}
and where we choose the conformal factor, $F$, to be the derivative of the function $f(R)$.  We find, using the transformation and equation (\ref{FE}), that for the $\tilde{g}$ system the following equation holds
\begin{eqnarray} \label{tran}
 \tilde{R}_{\alpha\beta} - \frac{1}{2}\tilde{R}\tilde{g}_{\alpha\beta} & = & \frac{1}{2}f(R)F^{-2}\tilde{g}_{\alpha\beta} - \frac{1}{2}F^{-1}R\tilde{g}_{\alpha\beta} + \frac{3}{2}F^{-2}\tilde{\nabla}_\alpha F \tilde{\nabla}_\beta F \nonumber \\
 & & - \frac{3}{4}F^{-2}\tilde{\nabla}^\gamma F\tilde{\nabla}_\gamma F \tilde{g}_{\alpha\beta}
\end{eqnarray}
where $\tilde{R}_{\alpha\beta}$ and $\tilde{R}$ are the Ricci tensor and scalar curvature built from the metric $\tilde{g}$ and where $\tilde{\nabla}_\alpha$ is covariant differentiation with respect to $\tilde{g}$.  If we then make the choices
\begin{equation} \label{phi}
 \phi = \sqrt{\frac{3}{16\pi}} \ln F
\end{equation}
and
\begin{equation} \label{V}
 V(\phi) = -\frac{1}{16\pi}\frac{f(R) - RF}{F^2}
\end{equation}
we find that this equation becomes
\begin{equation} \label{ESF}
 \tilde{R}_{\alpha\beta} - \frac{1}{2}\tilde{R}\tilde{g}_{\alpha\beta} = 8\pi \left( \tilde{\nabla}_\alpha \phi \tilde{\nabla}_\beta \phi - \frac{1}{2}\tilde{\nabla}_\gamma \phi \tilde{\nabla}^\gamma \phi \tilde{g}_{\alpha\beta} - V(\phi ) \tilde{g}_{\alpha\beta} \right)
\end{equation}
The right hand side is precisely the form for the energy-momentum tensor of a scalar field, $\phi$, with potential $V(\phi)$ and so we find that the conformal metric, $\tilde{g}$, is a solution to the Einstein equations with matter described by a scalar field.  

If we take the trace of the $f(R)$ field equations (\ref{FE}), we get
\begin{equation*}
 f'(R)R - 2f(R) + 3\square (f'(R)) = 0
\end{equation*}
or, since $F = f'(R)$, $$ FR - 2f(R) + 3\square F = 0 $$

Multiplying through by $\frac{1}{16\pi}\sqrt{\frac{16\pi}{3}}F^{-2}$ gives
\begin{equation} \label{trace}
 \frac{1}{16\pi}\sqrt{\frac{16\pi}{3}}\left[\frac{FR - 2f(R)}{F^2}\right] + \frac{3}{16\pi}\sqrt{\frac{16\pi}{3}}F^{-2}\square F = 0
\end{equation}
Recalling the definition of the potential, $V$, of the scalar field shows that the derivative, $V'(\phi)$, with respect to $\phi$ is exactly the 1st term on the left-hand-side.  We can then make the conformal transformation $g \mapsto \tilde{g}$ as before.  We get
\begin{equation*}
 \square F = F \tilde{\square}F - \tilde{\nabla}^\alpha F \tilde{\nabla}_\alpha F
\end{equation*}
Since $F$ and $\phi$ are related by $F = e^{\sqrt{\frac{16\pi}{3}}\phi}$, this then gives
\begin{eqnarray}
 \square F & = & \frac{16\pi}{3}F^2 \tilde{\nabla}^\gamma \phi \tilde{\nabla}_\gamma \phi + \sqrt{\frac{16\pi}{3}}F^2 \tilde{\square} \phi - \frac{16\pi}{3}F^2 \tilde{\nabla}^\alpha \phi \tilde{\nabla}_\alpha \phi \nonumber \\
 & = & \sqrt{\frac{16\pi}{3}}F^2 \tilde{\square} \phi \nonumber
\end{eqnarray}
We can then substitute this information into equation (\ref{trace}) above to show that
\begin{equation}
 V'(\phi) = \tilde{\square} \phi 
\end{equation}
which is exactly the equation of motion for the scalar field, $\phi$.

From the definition of the scalar field in terms of the function $f'(R)$ 
\begin{equation} \label{phif}
\phi = \sqrt{\frac{3}{16\pi}}\ln (f'(R))
\end{equation}
it is not clear whether we can express the scalar curvature, $R$, in terms of the scalar field, $\phi$, for all choices of $f(R)$.  We would like to be able to do this in order to express the function $V$ of $R$
\begin{equation} \label{V(R)}
 V(R) = -\frac{1}{16\pi}\frac{f(R) - Rf'(R)}{(f'(R))^2}
\end{equation}
 in terms of $\phi$, since $V$ represents the potential of the scalar field.  We also require that $V$ is smooth.  In order to achieve this, consider $\phi$ as a map $\phi : \mathbb{R} \rightarrow \mathbb{R}$ given by $$ R \mapsto \sqrt{\frac{3}{16\pi}}\ln (f'(R)) $$ In order to ensure that $R$ can be defined in terms of $\phi$, we require that the inverse of this map exists and is smooth.  By the inverse function theorem, we see that this is the case if $\phi'(R) \neq 0$ for all $R$.  So we must insist that $f''(R) \neq 0 $ for all $R$.  In other words $f'(R)$ is one-to-one and therefore either strictly increasing or strictly decreasing.  

Hence there is a nice correspondence between the vacuum $f(R)$ theory and the Einstein-scalar field system, where solutions of these two different theories are related by a conformal factor, $F$, which is the derivative of the function $f(R)$ which appears in the $f(R)$ theory.  

Therefore we can use known results on the dynamics of solutions of the Einstein-scalar field system to find results on the dynamics of solutions of the $f(R)$ theory.  For the Einstein-scalar field system, the dynamics of solutions depend on the form that the potential of the scalar field takes.  There are several known results for this system in the case of a homogeneous space-time, where different assumptions on the potential have been made.  In order to relate this to the $f(R)$ theory, we must translate these assumptions on the potential into assumptions on the function $f(R)$.  We will then gain results on the dynamics of the $f(R)$ theory which depend on the function $f(R)$.  The first case we will consider is the case where the potential admits a strictly positive minimum.

Since we will be interested in accelerated expansion, it is important to know the relationship between the Hubble parameter of the $f(R)$ system and that of the Einstein-scalar field system.  From the relationship between the metrics, we find that the scale factors are related by
\begin{equation}
 \tilde{a} = Fa
\end{equation}
and so the Hubble parameters are related by
\begin{equation} \label{Htilde}
 \tilde{H} = \frac{\dot{\tilde{a}}}{\tilde{a}} = \frac{F\dot{a} + \dot{F}a}{Fa} = \frac{\dot{a}}{a} + \frac{\dot{F}}{F} = H + \frac{\dot{F}}{F} 
\end{equation}

\subsection{The potential admits a strictly positive minimum}

Recall the relationship between the scalar field $\phi$ and the function $f(R)$ given in (\ref{phif}).  We see from this that we must have $f'(R) > 0$.  The potential $V(\phi)$ of the scalar field and the function $f(R)$ are related by equation (\ref{V(R)}).

We would like to consider cases of the scalar field where the potential is strictly positive and has a \emph{strictly positive minimum}, since these cases lead to accelerated expansion of the universe which does not stop \cite{Rendallaccexp}, and to study how the insistence of a strictly positive minimum gives rise to restrictions on $R, f(R)$ and the derivatives of $f(R)$.

To describe the asymptotic (late-time) behaviour of a solution of the $f(R)$ theory which satisfies these conditions, we follow the paper in \cite{Rendallposmin}, which gives results on the asymptotic behaviour of a solution to the Einstein-scalar field system, and apply these results to the equivalent $f(R)$ theory.

Consider first the case where the potential, $V$, is strictly positive.  This condition becomes, in terms of $R$ and $f(R)$,
\begin{equation} \label{V>0}
f(R) - Rf'(R) < 0
\end{equation}
for all $R$.

We can already prove some results on the $f(R)$ theory with this restriction in place.

We first prove an analogous theorem to Theorem 1 in \cite{Rendallposmin} for the scalar field system, which will be more useful for the translation to $f(R)$ theory.

\begin{theorem}
 Consider a solution, $\tilde{g}$, of the Einstein equations of Bianchi type I-VIII coupled to a nonlinear scalar field with potential $V$ of class $C^2$ defined on the (possibly infinite) interval $(\alpha,\beta)$ and satisfying the following two assumptions: 

1. $V(\phi) \geq V_0$ for a constant $V_0 > 0$ 

2. $V$ tends to $\infty$ as $\phi$ tends to $\alpha$ or $\beta$  \newline
If the solution is initially expanding ($\tilde{H} > 0$) and exists globally to the future, then for $t \rightarrow \infty$, the quantities $\tilde{R}$ and $\tilde{H}^2 - (8\pi/3)[\dot{\phi}^2 + V(\phi)]$ decay exponentially. The potential $V(\phi)$ converges to some positive constant $V_1$, $V'(\phi) \rightarrow 0$ and $\tilde{H} \rightarrow (8\pi V_1/3)^{1/2}$, where $\tilde{H}$ is the Hubble parameter for the Einstein-scalar field system. 
\end{theorem}

Note that this theorem will also hold in the presence of other matter satisfying the dominant and strong energy conditions, as in Theorem 1 in \cite{Rendallposmin}, but this is not needed for the purpose of this paper.

\begin{proof}
 Assumption 2 of the theorem above ($ V \rightarrow \infty$ as $\phi \rightarrow \alpha$ or $\beta$), together with the information that $V$ is bounded to the future (by conservation of energy), tells us that $\phi$ must remain bounded away from $\alpha$ and $\beta$ as $t \rightarrow \infty$.  Therefore, since $V$ is smooth, the derivative of $V$ with respect to $\phi$ must also be bounded to the future.  This information is sufficient to replace the second assumption in Theorem 1 of \cite{Rendallposmin}, since the proof only requires that $V'$ is bounded to the future.  The third assumption in Theorem 1 of \cite{Rendallposmin} is required to show that, if $\phi \rightarrow \pm \infty$ as $t \rightarrow \infty$, we know that $V'$ converges.  Now, since $V$ is defined on $(\alpha, \beta)$, then $\phi$ may only converge to $\pm \infty$ as $t \rightarrow \infty$ if one of $\alpha$ or $\beta$ is infinite.  However, we know from above that $\phi$ is bounded away from $\alpha$ and $\beta$ as $t \rightarrow \infty$ and so $\phi$ must not converge to $\pm \infty$ as $t \rightarrow \infty$.  Therefore, this assumption is no longer required.  With these alterations, the proof of Theorem 1 above now follows directly from the proof of Theorem 1 in \cite{Rendallposmin}.
\end{proof}

To translate this theorem to the $f(R)$ theory, we therefore need some conditions on the function $f(R)$ which will ensure that the potential blows up at its endpoints.  

We know already that the function $f'(R)$ is strictly positive and is one-to-one.  For the purpose of this paper, we will restrict ourselves to the case where $f'(R)$ is defined on the \emph{whole real line}.  The situation where $f'(R)$ has a smaller domain of definition can be treated in a similar way.  We assume for now that $f'(R)$ is strictly increasing.  Then, considering the graphs of $f'(R)$ against $R$ and $\phi$ against $f'(R)$, we find that both are strictly increasing.  Suppose that the range of $f'(R)$ is $(a,b)$ (the biggest range posssible being $(0, \infty)$).  Then $\phi$ can take values between $\sqrt{\frac{3}{16\pi}}\ln a$ and $\sqrt{\frac{3}{16\pi}}\ln b$.  So the domain of definition of the potential $V$ is $\left( \sqrt{\frac{3}{16\pi}}\ln a, \sqrt{\frac{3}{16\pi}}\ln b \right)$ (the biggest possible domain being the whole real line).  We are interested in how $V$ behaves as $\phi$ approaches $\sqrt{\frac{3}{16\pi}}\ln a$ and $\sqrt{\frac{3}{16\pi}}\ln b$ and we see that this is equivalent to studying the behaviour of $V$ as $R$ tends to $-\infty$ and $\infty$.

So we must consider how $V$ behaves as $R \rightarrow \pm \infty$ and show what assumptions must be made in order that $V$ blows up.  This will depend on the range of $f'(R)$.  

Consider the case where $f'(R)$ tends to a (positive) constant $a$ as $R \rightarrow -\infty$.  Then we can write, using the Landau notation \cite{Erdelyi}, $f'(R) = a + o(1)$ as $R \rightarrow -\infty$.  Integrating gives $f(R) = aR + o(R)$ as $R \rightarrow -\infty$ and multiplying $f'(R)$ by $R$ gives $Rf'(R) = aR + o(R)$.  Subtracting $Rf'(R)$ from $f(R)$ shows that the leading order terms cancel and we get $f(R) - Rf'(R) = o(R)$.  Recalling the definition of the potential in terms of the scalar curvature $R$, we find that as $R \rightarrow -\infty$,  $$ V = -\frac{1}{16\pi}\frac{o(R)}{(a + o(1))^2} = o(R) $$  This may or may not blow up, depending on the exact form of $f(R) - Rf'(R)$ as $R \rightarrow -\infty$.  In order to ensure that the potential tends to infinity as $R \rightarrow -\infty$, we require that $f(R) - Rf'(R) \rightarrow -\infty$ as $R \rightarrow -\infty$.

Now consider the case where $f'(R) \rightarrow 0$ as $R \rightarrow -\infty$.  In Landau notation, this means that $f'(R) = o(1)$ as $R \rightarrow -\infty$ and so $f(R) = o(R)$ as $R \rightarrow -\infty$.  So we find that $Rf'(R) = o(R)$ as $R \rightarrow -\infty$ also.  Since we know that $f'(R)$ tends to zero as $R$ tends to negative infinity, we see that the denominator of the potential tends to zero.  Therefore, provided the numerator does not tend to zero faster than the denominator, we will find that the potential blows up as $R$ tends to negative infinity.  One way to ensure that this is the case is to restrict the function $f(R)$ so that the expression $f(R) - Rf'(R)$ is bounded away from zero.  (It should be noted that this is not a necessary condition, but is a sufficient one.)

This covers the situation as $R \rightarrow -\infty$.  Now let us consider what happens as $R \rightarrow \infty$.  

First, suppose that the endpoint of the range of $f'(R)$ is finite and equal to $b$.  (So $f'(R) \rightarrow b$ as $R \rightarrow \infty$).  Then $f'(R) = b + o(1)$ for $R \rightarrow \infty$.  Integrating gives $f(R) = bR + o(R)$ and we also have $Rf'(R) = bR + o(R)$ and $(f'(R))^2 = (b + o(1))^2$ as $R \rightarrow \infty$.  Therefore we have the same situation as in the case for $f'(R) \rightarrow a$ as $R \rightarrow -\infty$.  Once again, we find that the potential will blow up provided $f(R) - Rf'(R) \rightarrow -\infty$ as $R \rightarrow \infty$.

Now suppose that $f'(R)$ tends to infinity as $R$ tends to infinity.  The situation is now more complicated.  If $f'(R) \rightarrow \infty$, then $f(R)$ must also tend to infinity.  So we have the situation that $R, f(R)$ and $f'(R)$ all tend to infinity.  We know that the denominator of the potential tends to infinity as $R \rightarrow \infty$.  So we must insist that the numerator tends to negative infinity as $R \rightarrow \infty$.  However, this is not enough to ensure that the potential blows up.  We must also insist that the numerator tends to negative infinity \emph{faster} than the denominator tends to infinity.  This requires that $(f'(R))^2 = o(f(R) - Rf'(R))$.

The case where $f'(R)$ is strictly decreasing can be treated in a similar way and we find the following restrictions on $R, f(R)$ and $f'(R)$.

\textbf{(F1)} If $f'(R) \rightarrow a$ as $R \rightarrow \infty$ or $-\infty$, we require $f(R) - Rf'(R) \rightarrow -\infty$ as $R \rightarrow \infty$ or $-\infty$.

\textbf{(F2)} If $f'(R) \rightarrow 0$ as $R \rightarrow \infty$ or $-\infty$, we require that $f(R) - Rf'(R)$ is bounded away from zero as $R \rightarrow \infty$ or $-\infty$

\textbf{(F3)} If $f'(R) \rightarrow \infty$ as $R \rightarrow \infty$ or $-\infty$, we require that $f(R) - Rf'(R) \rightarrow -\infty$ as $R \rightarrow \infty$ \emph{and} that $(f'(R))^2 = o(Rf'(R) - f(R))$.

We are now ready to state the theorem for the $f(R)$ theory.

\begin{theorem}
 Consider a solution, $g$, to the $f(R)$ field equations in a Bianchi type I-VIII space-time, where $f(R)$ satisfies the following properties:
\begin{enumerate}
 \item $f(R) - Rf'(R) < 0$ for all $f'(R) > 0$
 \item $f(R) - Rf'(R)$ satisfies the three restrictions (F1), (F2), (F3)
\end{enumerate}
If the solution satisfies $H + \frac{\dot{F}}{F} > 0$ initially and exists globally to the future, then for $t \rightarrow \infty$, the quantity $-\frac{f(R) - Rf'(R)}{(f'(R))^2}$ tends to some positive constant $K$, $\frac{2f(R) - Rf'(R)}{(f'(R))^2}$ tends to zero and $H$ tends to $\left(\frac{K}{6}\right)^{1/2}$, where $H$ is the Hubble parameter.

\end{theorem}

\begin{proof}
 From the relationships between the function $f(R)$ of the $f(R)$ theory and the scalar field $\phi$ of the Einstein-scalar field system, between $f(R)$ and the potential, $V(\phi)$, and between the Hubble parameters of the two systems, we see from the discussion above that, if $g$ satisfies the assumptions of this theorem, then $\tilde{g} = Fg$ is a solution of the Einstein-scalar field equations satisfying the assumptions of Theorem 1 above.  Therefore, we can use the results of the latter to show that, if the solution is initially expanding ($\tilde{H} > 0$) and exists globally to the future, then for $t \rightarrow \infty$, the potential $V(\phi)$ converges to some constant $V_1$, $V'(\phi) \rightarrow 0$ and the Hubble parameter, $\tilde{H}$, of the Einstein-scalar field system tends to $ \left( \frac{8\pi V_1}{3}\right)^{1/2}$.  

Translating this back to the $f(R)$ system, using the relationships between $f(R)$ and $\phi$ and between $f(R)$ and $V(\phi)$, gives the first two results of Theorem 2 above, where the constant $K$ is equal to $16\pi V_1$ (and is therefore positive, since $V_1$ is positive).  Recall that the Hubble parameters are related by $$ \tilde{H} = H + \frac{\dot{F}}{F} $$
We can use the information from the proof of Theorem 1, for which we refer the reader to the proof of Theorem 1 in \cite{Rendallposmin}, that $\dot{\phi} \rightarrow 0$ as $t \rightarrow \infty$.  From the definition of $\phi$ in terms of $F$ (equation (\ref{phi})), we get
\begin{equation} \label{dotphi}
 \dot{\phi} = \sqrt{\frac{3}{16\pi}} \frac{\dot{F}}{F}
\end{equation}
and so we find that $\frac{\dot{F}}{F} \rightarrow 0$ as $t \rightarrow \infty$ and hence that the Hubble parameter $H$ is equal to $\tilde{H}$ at late times.  This concludes the proof of Theorem 2.
\end{proof}

The theorem therefore states that, as $t \rightarrow \infty$, the quantity $ \frac{2f(R) - Rf'(R)}{(f'(R))^2} $ tends to zero and the quantity $-\frac{f(R) - Rf'(R)}{(f'(R))^2}$ tends to a constant $K$.  Suppose now that any points where the quantity $\frac{2f(R) - Rf'(R)}{(f'(R))^2}$ equals zero are \emph{isolated}.  Then this, together with the information that this quantity tends to zero as $t \rightarrow \infty$, tells us that $R$ must \emph{converge} as $t \rightarrow \infty$.  But $-\frac{f(R) - Rf'(R)}{(f'(R))^2} \rightarrow \infty$ as $R \rightarrow \pm \infty$ (from the assumptions (F1), (F2), (F3) of Theorem 2), together with $-\frac{f(R) - Rf'(R)}{(f'(R))^2} \rightarrow K$ as $t \rightarrow \infty$, shows that $R$ must converge to a \emph{finite} limit, $R_c$ say, as $t \rightarrow \infty$.  So we find that as $t \rightarrow \infty$, $R \rightarrow R_c$ with $2f(R_c) - R_cf'(R_c) = 0$.  

Considering again the equation for $V$ in terms of $f(R)$, given in (\ref{V(R)}) and taking its derivative with respect to $\phi$, tells us that 
\begin{equation} \label{potderiv}
 \frac{dV}{d\phi} = \frac{1}{16\pi}\sqrt{\frac{16\pi}{3}}\left[ \frac{2f(R) - Rf'(R)}{\left(f'(R)\right)^2}\right]
\end{equation}
We therefore see that the limit, $R_c$, of $R$ as $t \rightarrow \infty$ corresponds to a critical point of the potential, $V$.  Furthermore we can use the fact that $2f(R_c) = Rf'(R_c)$ to find the value of $V$ at this point, and therefore the constant $K$ of Theorem 2, in terms of $R_c$ .  We find that 
\begin{equation*}
 K = \frac{R_c^2}{4f(R_c)}
\end{equation*}
which shows that $H$ tends to a positive limit and so we have accelerated expansion at late times for the solution of the $f(R)$ theory.  

To gain more detailed information about the late-time asymptotics of the solution, we impose further conditions on the potential, $V$.  We assume now that $V$ tends to a strictly positive \emph{non-degenerate minimum} as $t \rightarrow \infty$.   To discover how this restricts the function $f(R)$, we use (\ref{phif}) and (\ref{V(R)}) to find the second derivative of the potential with respect to $\phi$.  We get
\begin{equation} \label{2ndderiv}
 \frac{d^2V}{d\phi^2} = \frac{1}{3}\left[\frac{(f'(R))^2 + Rf'(R)f''(R) - 4f(R)f''(R)}{(f'(R))^2(f''(R))}\right]
\end{equation}
where $f''(R)$ denotes the second derivative of $f$ with respect to $R$.
The point $R_c$ corresponds to the critical point of the potential.  Recalling that this point must satisfy $2f(R_c) = R_c f'(R_c)$ and substituting this into (\ref{2ndderiv}), we get
\begin{equation*} 
 \frac{d^2V}{d\phi^2} = \frac{1}{3}\left[\frac{1}{f''(R_c)} - \frac{R_c}{f'(R_c)}\right]
\end{equation*}
Then, in order that the critical point is a \emph{non-degenerate minimum}, we require that the second derivative is positive.  This requires that the term in brackets is positive.  Since the expression $f(R_c) - R_c f'(R_c)$ must be negative (from (\ref{V>0})) and we have the equality $2f(R_c) = Rf'(R_c)$, we find that $R_c$ and $f(R_c)$ must both be positive.  Then in order that the term in brackets is positive, we need
\begin{equation*}
 f''(R_c) > 0
\end{equation*}
and
\begin{equation*}
 f''(R_c) < \frac{f'(R_c)}{R_c}
\end{equation*}
So, in order that the critical point is a non-degenerate minimum, we must have that the second derivative of $f$ with respect to $R$ at this point lies in the interval $$\left(0, \frac{f'(R_c)}{R_c}\right)$$

We can now use Theorem 2 in \cite{Rendallposmin} to get more detailed information on the asymptotics in the future.

\begin{theorem}
 Let $g$ be a solution to the $f(R)$ field equations satisfying the hypotheses of Theorem 2.

Then, if $R_c$ is the limit of $R$ as $t \rightarrow \infty$, and $$ 0 < f''(R_c) < \frac{f'(R_c)}{R_c} $$ we find that the quantities $f'(R_c), -\frac{f(R_c) - R_cf'(R_c)}{(f'(R_c))^2} - K$ and $H - H_1$ decay exponentially as $t \rightarrow \infty$, where $H_1 = \sqrt{\frac{K}{6}}$.  We also find that $f'(R_c)e^{-2H_1t}g_{\alpha\beta}$ converges to a limit.
\end{theorem}

\begin{proof}
 The assumption made on the second derivative of $f(R)$ as $t \rightarrow \infty$ is equivalent to the assumption made on the second derivative of the potential in Theorem 2 in \cite{Rendallposmin}.  It is then seen that the conformally transformed metric, $\tilde{g}_{\alpha\beta} = Fg_{\alpha\beta}$, will satisfy all the assumptions required for that theorem and hence that the conclusions will follow for the metric $\tilde{g}_{\alpha\beta}$.  Using the information gained in Theorem 2 of this paper and the discussion after, we see that the results of Theorem 2 in \cite{Rendallposmin} can be transferred onto the metric, $g$, of the $f(R)$ theory to give the conclusions of Theorem 3 above.
\end{proof}

\begin{rmk}
 It should perhaps be noted here that the condition that the minimum of the potential is \emph{strictly positive} is very important for the accelerated expansion result.  To see this, consider a function of the form $$ f(R) = R + \alpha R^2 $$ which gives a potential $$ V(\phi) = \frac{1}{48\pi \alpha} \left( 1 - e^{-\sqrt{16\pi/3}\phi} \right)^2 $$ that admits a zero minimum at $\phi = 0$.  The paper \cite{Miritzis} has studied flat and open FLRW models with a scalar field that has a potential with a local zero minimum.  It is shown that the solution is ever expanding, but that $\phi$ and $H$ asymptotically approach zero and therefore the solution does not undergo accelerated expansion at late times.  Therefore, as discussed in section 4 of \cite{Miritzis2}, the corresponding $f(R)$ function which gives rise to an Einstein-scalar field solution with this type of potential cannot provide a mechanism for late accelerating expansion of the universe.
\end{rmk}

To finish, I would like to bring attention to an example \cite{Appleby} in the literature of a function $f(R)$ that fits the criteria of this section and is therefore a solution of the $f(R)$ theory that undergoes accelerated expansion.

In the paper, the function studied is 
\begin{equation}
 f(R) = \frac{1}{2}R + \frac{1}{2a}\log{\left(\cosh(aR)-\tanh(b)\sinh(aR)\right)}
\end{equation}
where restrictions are placed on the parameters $a$ and $b$ by requiring $a>0$ and $b \gtrsim 1.2$.  For the purpose of this paper, we assume $a > 0$, but will not place restrictions on $b$ until they are needed.  The derivative of $f(R)$ is given by
\begin{equation}
 f'(R) = \frac{1}{2}\left[ 1 + \tanh(aR - b)\right]
\end{equation}
and this clearly satisfies the condition that $f'(R) > 0$ for all $R$.  Note that the range of $f'(R)$ is $(0,1)$ and hence, from (\ref{phif}), that $\phi$ is only defined for $(-\infty, 0)$.  This means that the domain of definition for the potential is $(-\infty, 0)$.  

The condition that the potential is strictly positive, given by equation (\ref{V>0}), becomes
\begin{equation*}
 \frac{1}{2a}\log{\left( \cosh(aR) - \tanh(b)\sinh(aR)\right)} - R\tanh(aR - b) < 0
\end{equation*}
The expression on the left-hand-side tends to $- \infty$ as $R \rightarrow \pm \infty$ and contains only one turning point, a maximum, which takes the value zero at $R = 0$.  So we find that the inequality is true for all $R$, except the point $R=0$.  

Now, the function $f'(R)$ behaves in the following way as $R \rightarrow \pm \infty$.  As $R \rightarrow -\infty , f'(R) \rightarrow 0$ and as $R \rightarrow \infty, f'(R) \rightarrow 1$.  So in order that the second assumption in Theorem 2 (corresponding to the blow-up of the potential at its endpoints) holds, we require that $f(R) - Rf'(R) \rightarrow -\infty$ as $R \rightarrow \pm \infty$.  But we have already shown that this is the case, and so the function satisfies the second condition of Theorem 2.

The fact that we do not have that $f(R) - Rf'(R) < 0$ for all $R$ seems to cause some problems.  However, we can get round this by considering the conservation of energy, which tells us that if a solution falls into a strictly positive local minimum of the potential, it will get trapped there. If this is the case then we are free to change the shape of the potential away from this local minimum without affecting the solution.  Hence we need not worry about the behaviour of the potential away from the strictly positive local minimum.


The discussion following Theorem 2 then tells us that, as $t \rightarrow \infty$, $R \rightarrow R_c$ with $2f(R_c) = Rf'(R_c)$.  We can apply this to the function under consideration to find the possibilities for the limit, $R_c$, of $R$ as $t \rightarrow \infty$.  So let us study the zeroes of the function $Q(R) = 2f(R) - Rf'(R)$. We get
\begin{equation}
 Q(R) = \frac{R}{2} + \frac{1}{a}\log\left(\frac{\cosh(aR-b)}{\cosh(b)}\right) - \frac{R}{2}\tanh (aR-b)
\end{equation}
Consider what happens to $Q(R)$ as $R$ tends to plus and minus infinity.  As $R \rightarrow -\infty$, $\cosh (aR-b) - \frac{1}{2}e^{-(aR-b)} \rightarrow 0$ and  $\tanh (aR-b) \rightarrow -1$, so $$ Q(R) \rightarrow -\frac{1}{a}\log 2 + \frac{b}{a} - \frac{1}{a}\log (\cosh (b)) $$  This expression is negative for any choice of $b$, since $a>0$.  To see this, consider the graph of this expression against $b$.  We see that it tends to $-\infty$ as $b$ tends to $-\infty$ and zero as $b$ tends to $\infty$ and has no turning points.  Therefore, as $R$ tends to $-\infty$, $Q(R)$ tends to a negative limit.  As $R$ tends to plus infinity, $\cosh (aR-b) - \frac{1}{2}e^{aR-b} \rightarrow 0$ and $\tanh (aR-b) \rightarrow 1$, so we find that $Q(R)$ tends to infinity.  

Now, at $R=0$ we see that $Q=0$, which gives one possible choice for $R_c$.  We can also see that the function $Q$ is increasing between minus infinity and zero, so $Q$ does not have any zeroes for $R<0$.  We would like to know if there are any zeroes for positive $R$.  To investigate further, consider the second derivative of $Q$, which is given by $$ Q''(R) = \frac{Ra^2\tanh (aR-b)}{\cosh^2(aR-b)} $$  For $b<0$, the second derivative is always positive and so the function $Q(R)$ is convex and will clearly have no more zeroes.  In this case the only possibility for $R_c$ is $R_c = 0$.  But if this is the case then we find that the constant $K$ in Theorem 2 is also zero and this does not give rise to a solution which undergoes accelerated expansion.  So if we want a solution that gives accelerated expansion, we require more options for $R_c$ and hence a zero of $Q(R)$ at some $R > 0$, where the constant $K$ of Theorem 2 will then be positive.  Therefore we must assume $b>0$. Clearly, at $R = \frac{b}{a}$, we have $$ Q''(R) = 0 $$  This tells us that the graph changes shape at $\frac{b}{a}$.  For $R<\frac{b}{a}$ the second derivative is negative and for $R>\frac{b}{a}$ it is positive.  Now, at $R = \frac{b}{a}$, the function $Q$ is given by $$ Q(R) = \frac{1}{2a}\left( b - 2\log (\cosh (b))\right) $$  Clearly if the value of $Q(R)$ is negative at $R = \frac{b}{a}$, then there must be 2 more zeroes of the function $Q$ (since at $R=0$, the function $Q$ is zero and increasing and as $R\rightarrow \infty$, $Q\rightarrow \infty$).   Since $a>0$, we must investigate for what values of $b$ we have $$ b - 2\log (\cosh (b)) < 0 $$  

It can be shown that choosing $b \geq 2 \log 2$ ensures that this inequality holds.  To see this, note that if $$ b - 2\log (\cosh (b)) < 0 $$ then $$ e^{b} < e^{\log (\cosh^2(b))} $$ and so $$ 2e^{\frac{3b}{2}} - e^{2b} - 1 < 0 $$  Making the substitution $x = e^{b/2}$ gives $$ 2e^{\frac{3b}{2}} - e^{2b} - 1 = 2x^3 - x^4 - 1 $$ the right-hand-side of which can be factorised as $$ 2x^3 - x^4 - 1 = (x-1)(-x^3 + x^2 + x + 1)$$  Since we have assumed that $b>0$ we see that $x > 1$ and so $x-1 > 0$ hence we see that $$ b - 2\log (\cosh (b)) < 0 \Leftrightarrow -x^3 + x^2 + x + 1 < 0 $$ and since $x=e^{b/2}$ we must have positive $x$.  Studying the polynomial $-x^3 + x^2 + x + 1$ shows that its positive root lies between $x=1$ where the polynomial is positive and $x=2$ where the polynomial is negative.  Hence, if we take $x \geq 2$, then we are sure that the polynomial is negative.  This corresponds to $b \geq 2\log 2$. 

Therefore, if $b \geq 2\log 2$, then the function $Q(R)$ has 2 further zeroes and so we have another 2 possibilities for the limit, $R_c$, of $R$ as $t \rightarrow \infty$ (since we have already ruled out the case $R_c = 0$).

For Theorem 3, we require an extra condition on the $f(R)$ function, that $$ 0 < f''(R_c) < \frac{f'(R_c)}{R_c} $$  We see that this is equivalent to $Q'(R) > 0$, since from the definition of $Q(R)$ we get $$ Q'(R) = f'(R) - Rf''(R) $$  Using all the information we have gained about $Q(R)$, we can see that the third zero of $Q$ must have positive gradient and the second must have negative gradient.  (To see this, consider the sign of $Q'(R)$ immediately before and after $R = \frac{b}{a}$).  Therefore, if the limit of $R$ as $t \rightarrow \infty$ is the third possibility given, we see that this $R_c$ satisfies the conditions of Theorem 3 and so this choice of $f(R)$ admits a solution of the $f(R)$ theory which gives rise to the accelerated expansion described in Theorem 3.  

To give a concrete example, we choose, as in the plots given in \cite{Appleby}, $a = 2$ and $b = 1.5$.  If we consider the graph of the function $2f(R) - Rf'(R)$, we see that $2f(R) - Rf'(R)$ has 3 zeroes at $R = 0, R \simeq 0.4818758472$ and at $R \simeq 1.385168809$.  So we have 3 candidates for the choice of $R_c$.  The first of these gives $K = 0$ in Theorem 2 and so does not give rise to accelerated expansion.  In the second and third cases, we get $$ f''(R) = \frac{1}{\cosh^2(2R - 1.5)} $$ and $$\frac{f'(R)}{R} = \frac{1}{2R}\left(1 + \tanh(2R - 1.5)\right) $$  We find that the second case gives $$ f''(R) > \frac{f'(R)}{R} $$ (which does not satisfy the assumptions of Theorem 3) but the third case gives $$ f''(R) < \frac{f'(R)}{R} $$  Therefore this case satisfies the assumptions of Theorem 3, as required.  See the paper \cite{Appleby} for a plot of the potential.  

\subsection{Intermediate inflation}

We have considered the case where the potential is strictly positive and has a minimum.  The next step is to allow the potential to tend to zero as the scalar field tends to infinity.  In particular, we would like to consider potentials which satisfy the following two conditions:

1. $V(\phi) > 0$ with $V(\phi) \rightarrow 0$ as $\phi \rightarrow \infty$

2. $V'(\phi) < 0$ \newline
The paper \cite{slowroll} gives some results on the dynamics of scalar fields which satisfy these conditions.  It is known that accelerated expansion will occur provided the potential does not decay too rapidly.  Therefore it is necessary to restrict the potential further, e.g. placing a restriction on the ratio $V'/V$ will ensure accelerated expansion at late times.  The paper \cite{slowroll} places the following restriction on $V'/V$:

3. $V'(\phi)/V(\phi) \rightarrow 0$ as $\phi \rightarrow \infty$ \newline
Then Theorem 1 of \cite{slowroll} shows that the solution undergoes accelerated expansion at late times.  
We can use this theorem to prove results on the late-time dynamics of the $f(R)$ theory.

To proceed, note that the assumptions 1, 2 and 3 above on the potential of the scalar field can be translated into assumptions on the function $f(R)$ by means of equations (\ref{V(R)}) and (\ref{potderiv}).  We get

\vspace{.1in}
(f1) $f(R) - Rf'(R) < 0$ with $\frac{f(R) - Rf'(R)}{(f'(R))^2} \rightarrow 0$ as $f'(R) \rightarrow \infty$

\vspace{.1in}
(f2) $2f(R) - Rf'(R) < 0$ for all $R$

\vspace{.1in}
(f3) $\frac{Rf'(R)}{f(R)} \rightarrow 2$ as $f'(R) \rightarrow \infty$.  \vspace{.1in} \newline
We assume (analogously to the paper \cite{slowroll}) that we are in a situation where $f'(R) \rightarrow \infty$ as $t \rightarrow \infty$.  This assumption will be required for the rest of this section.  We can now use Theorem 1 of \cite{slowroll} to prove the following result on solutions of the $f(R)$ theory that satisfy conditions (f1), (f2) and (f3).  

\begin{theorem}
 Consider a spatially flat, homogeneous and isotropic solution of the $f(R)$ field equations satisfying conditions (f1), (f2) and (f3) above.  Suppose that the solution satisfies $H > 0$ and $\frac{\dot{F}}{F} > 0$ initially, where $F = f'(R)$ and $H$ is the Hubble parameter.  If the solution exists globally to the future, then the scale factor, $a$, satisfies $\ddot{a} > 0$ for $t$ sufficiently large and the solution undergoes accelerated expansion at late times.
\end{theorem}

\begin{proof}
 Firstly, we have shown that, if $g$ is a solution to the $f(R)$ field equations, given in (\ref{FE}), satisfying conditions (f1),(f2) and (f3), then $\tilde{g} = Fg$ is a solution to the Einstein-scalar field equations, given in (\ref{ESF}), where the potential of the scalar field satisfies conditions 1 - 3 above.  Now, recall that the relationship between the Hubble parameter, $\tilde{H}$, of the Einstein-scalar field system and the Hubble parameter, $H$, of the $f(R)$ theory is given by (\ref{Htilde}) $$ \tilde{H} = H + \frac{\dot{F}}{F} $$ and that the scalar field and the function $F$ are related by (\ref{phif}) $$ \phi = \sqrt{\frac{3}{16\pi}}\ln F $$
Therefore $\dot{\phi} = \sqrt{\frac{3}{16\pi}}\frac{\dot{F}}{F}$ and we find that the assumptions in the theorem above that $H > 0$ and $\frac{\dot{F}}{F} > 0$ initially, tell us that $\tilde{H} > 0$ initially and that $\dot{\phi} > 0$ initially, which ensures that all assumptions of Theorem 1 in \cite{slowroll} are satisfied.  Therefore the results of that theorem hold for the solution, $\tilde{g}$, of the Einstein-scalar field system.  In particular, $\ddot{\tilde{a}} > 0$ for sufficiently large $t$ and the solution undergoes accelerated expansion.  

We would now like to show that the scale factor, $a$, of the $f(R)$ theory also satisfies $\ddot{a} > 0$ for sufficiently large $t$.  To do this, we note that this is equivalent to proving $\dot{H} + H^2 > 0$.  We know from Theorem 1 of \cite{slowroll} that the inequality $\dot{\tilde{H}} + \tilde{H}^2 > 0$ (which is equivalent to $\frac{3\tilde{H}^2}{8\pi V} < 3/2$) holds for sufficiently large $t$.  So we consider the relationship between $H$ and $\tilde{H}$.  We know already that $\tilde{H} = H + \frac{\dot{F}}{F}$ and that $\dot{\phi} = \sqrt{\frac{3}{16\pi}}\frac{\dot{F}}{F}$.  During the proof of Theorem 1 in \cite{slowroll}, it is noted that $$ \frac{\dot{\phi}}{\tilde{H}} \rightarrow 0 \mbox{ as } \frac{3\tilde{H}^2}{8\pi V} \rightarrow 1 \mbox{ and } \frac{3\tilde{H}^2}{8\pi V} \rightarrow 1 \mbox{ as } t \rightarrow \infty $$ So we find that $$ \frac{\dot{\phi}}{\tilde{H}} \rightarrow 0 \mbox{ as } t \rightarrow \infty $$ But from the first assumption on the potential, we know that $V \rightarrow 0$ as $t \rightarrow \infty$ and so, since $\frac{3\tilde{H}^2}{8\pi V} \rightarrow 1$ as $t \rightarrow \infty$, this means that $\tilde{H} \rightarrow 0$ as $t \rightarrow \infty$.  Hence, from above, we must also have that $\dot{\phi} \rightarrow 0$ as $t \rightarrow \infty$.  

Therefore we find that $\frac{\dot{F}}{F} \rightarrow 0$ as $t \rightarrow \infty$.  So, from the relationship between $H$ and $\tilde{H}$, we see that $H$ behaves like $\tilde{H}$ at late times.  Now consider $\dot{H}$.  This is related to the $\dot{\tilde{H}}$ of the Einstein-scalar field system by $$ \dot{\tilde{H}} = \dot{H} + \frac{\ddot{F}}{F} - \left(\frac{\dot{F}}{F}\right)^2$$  We know already that the third term tends to zero as $t$ tends to infinity.  From the relationship between $F$ and $\phi$ we find that $$ \frac{\ddot{F}}{F} = \frac{16\pi}{3} \dot{\phi}^2 + \sqrt{\frac{16\pi}{3}}\ddot{\phi} $$ Once again, we know that $\dot{\phi} \rightarrow 0$ as $t \rightarrow \infty$.  We also know, from \cite{slowroll}, that the following equation holds for the Einstein-scalar field system:
\begin{equation}
 \ddot{\phi} = -3\tilde{H}\dot{\phi} - V'(\phi)
\end{equation}
and that $\tilde{H} \rightarrow 0$ and $\dot{\phi} \rightarrow 0$ as $t \rightarrow \infty$.  From the first and third assumptions on the potential of the scalar field, we see that $V \rightarrow 0$ and $V'/V \rightarrow 0$ as $t \rightarrow \infty$, which tells us that $V' \rightarrow 0$ as $t \rightarrow \infty$.  So we find that $\ddot{\phi}$ must also tend to zero as $t$ tends to infinity.  Thus we find that $$ \dot{\tilde{H}} - \dot{H} \rightarrow 0 \mbox{ as } t \rightarrow \infty $$

Therefore the behaviour of $\dot{H} + H^2$ as $t \rightarrow \infty$ mimics that of $\dot{\tilde{H}} + \tilde{H}^2$ and, in particular, $\dot{H} + H^2 > 0$ at late times.  This concludes the proof.

\end{proof}

We would now like to give an example of a function $f(R)$ that satisfies the conditions (f1), (f2) and (f3) above.  To find such an example, we note that the function $$ f(R) = R^2 $$ gives rise to a scalar field with \emph{constant} potential, whereas the function $$ f(R) = R^n $$ with $ n > 2$ gives a scalar field with \emph{exponential} potential.  We are looking for something that lies in between these two functions.  Consider $$ f(R) = R^2 \log R $$ defined on $(e^{-1/2}, \infty)$ so that $f'(R) > 0$. Then $f(R) > 0$ with $f'(R) \rightarrow \infty$ as $R \rightarrow \infty$.  Now for the first condition, (f1), we get $$f(R) - Rf'(R) = - R^2 \log R - R^2$$ which is negative for all $R$ in the domain and $$\frac{f(R) - Rf'(R)}{(f'(R))^2} = \frac{-1 - \frac{1}{\log R}}{(2(\log R)^{1/2} + (\log R)^{-1/2})^2} \rightarrow 0 \mbox{ as } f'(R) \rightarrow \infty$$ For the second condition, (f2), $$ 2f(R) - Rf'(R) = -R^2 $$ which is negative for all $R$ in the domain and for the third condition, (f3), $$ \frac{Rf'(R)}{f(R)} = 2 + \frac{1}{\log R} \rightarrow 2 \mbox{ as } R \rightarrow \infty$$ as required.  

Therefore, this particular choice of $f(R)$ satisfies the conditions of Theorem 4 and is another example of an $f(R)$ theory which undergoes accelerated expansion at late-times.

In the paper \cite{slowroll}, it was shown that the third condition on the potential that $V'/V \rightarrow 0$ as $\phi \rightarrow \infty$ can be relaxed to $$ \lim \sup \left(-V'/V\right) < 4\sqrt{\pi} $$ without altering the result that $\ddot{\tilde{a}} > 0$.  This then gives accelerated expansion for a more general type of potential.  In particular, this includes the case of the exponential potential.

We can similarly relax the condition (f3) on the $f(R)$ theory to prove accelerated expansion results on a more general type of $f(R)$ function.  Recall that 
\begin{eqnarray}
  \frac{-V'}{V} & = & \sqrt{\frac{16\pi}{3}}\left(\frac{2f(R)-Rf'(R)}{f(R)-Rf'(R)}\right) \nonumber \\
 & = & \sqrt{\frac{16\pi}{3}}\left(\frac{f(R)}{f(R)-Rf'(R)} + 1\right) \nonumber
\end{eqnarray}
Then our alternative condition for the $f(R)$ theory is $$\lim \sup \left(\frac{f(R)}{f(R) - Rf'(R)}\right) < \sqrt{3} - 1$$

We can now prove the following theorem:

\begin{theorem}
 Consider a spatially flat homogeneous and isotropic solution to the $f(R)$ field equations satisfying conditions (f1) and (f2) above with $$ \lim \sup \left(\frac{f(R)}{f(R) - Rf'(R)}\right) < \sqrt{3} - 1 $$  Suppose that the solution satisfies $H > 0$ and $\frac{\dot{F}}{F} > 0$ initially, where $F = f'(R)$ and $H$ is the Hubble parameter.  If the solution exists globally to the future, then the scale factor, $a$, satisfies $\ddot{a} > 0$ for $t$ sufficiently large and the solution undergoes accelerated expansion at late times.
\end{theorem}

\begin{proof}
 As in the proof of the previous theorem, we know that if $g$ is a solution to the $f(R)$ field equations, given in (\ref{FE}), satisfying the conditions of Theorem 5, then $\tilde{g} = Fg$ is a solution to the Einstein-scalar field equations, given in (\ref{ESF}), where the potential of the scalar field satisfies conditions 1 and 2 above on the potential, together with the condition $\lim \sup (-V'/V) < 4\sqrt{\pi}$.  We also know, as in the previous theorem, that, if $H > 0$ and $\frac{\dot{F}}{F} > 0$ initially, then $\tilde{H} > 0$ and $\dot{\phi} > 0$ initially.  Therefore, the solution $\tilde{g}$ of the Einstein-scalar field system satisfies the assumptions of Theorem 2 in \cite{slowroll} and we know that, for this solution, $\lim \sup \left(\frac{3\tilde{H}^2}{8\pi V}\right) < 3/2$, which implies $\dot{\tilde{H}} + \tilde{H}^2 > 0$, which gives $\ddot{\tilde{a}} > 0$.

Now recall that $$ \tilde{H} = H + \frac{\dot{F}}{F} $$ and $$ \dot{\tilde{H}} = \dot{H} + \frac{\ddot{F}}{F} - \left(\frac{\dot{F}}{F}\right)^2 $$  By the definition of the scalar field in terms of the function $F = f'(R)$, we have $$ \dot{\phi} = \sqrt{\frac{3}{16\pi}} \frac{\dot{F}}{F}$$  In the proof of Theorem 2 in \cite{slowroll}, we find that $$ \lim \sup \left(\frac{3\tilde{H}^2}{8\pi V}\right) < \frac{3}{2} $$ as $t \rightarrow \infty$.  Then, since $$ \frac{\dot{\phi}}{\tilde{H}} = \sqrt{\frac{3}{4\pi} - \frac{2V}{\tilde{H}^2}} $$ we find that $$\lim \sup \left(\frac{\dot{\phi}}{\tilde{H}}\right) < \sqrt{\frac{1}{2\pi}} $$ as $t \rightarrow \infty$.  Now $ \lim \sup \left(\frac{3\tilde{H}^2}{8\pi V}\right) < \frac{3}{2}$ as $t \rightarrow \infty$ together with $V \rightarrow 0$ as $t \rightarrow \infty$ (from the first assumption of Theorem 2 in \cite{slowroll}) implies that $$ \tilde{H} \rightarrow 0 \mbox{ as } t \rightarrow \infty$$ Then, this information together with $\lim \sup \left(\frac{\dot{\phi}}{\tilde{H}}\right) < \sqrt{\frac{1}{2\pi}}$ as $ t \rightarrow \infty$ implies that $$ \dot{\phi} \rightarrow 0 \mbox{ as } t \rightarrow \infty$$  Hence we know that $$ \frac{\dot{F}}{F} \rightarrow 0 \mbox{ as } t \rightarrow \infty $$ As before, we know that $\ddot{\phi} = -3\tilde{H}\phi - V'(\phi)$ and that $\dot{\phi} \rightarrow 0 $ and $\tilde{H} \rightarrow 0$ as $t \rightarrow \infty$.  We also know, from the assumptions of Theorem 2 in \cite{slowroll} that $$ \lim \sup \left(-V'/V\right) < 4\sqrt{\pi} \mbox{ as } t \rightarrow \infty$$ and that $V \rightarrow 0$ as $t \rightarrow \infty$.  Therefore we must also have that $V' \rightarrow 0$ as $t \rightarrow \infty$.  Hence, from above, $$\ddot{\phi} \rightarrow 0 \mbox{ as } t \rightarrow \infty$$

In summary, we have shown that $\frac{\dot{F}}{F} \rightarrow 0$ and $\frac{\ddot{F}}{F} \rightarrow 0$ as $ t \rightarrow \infty$ and hence that $\dot{H} + H^2$ behaves like $\dot{\tilde{H}} + \tilde{H}^2$ for large $t$ and in particular that $\ddot{a} > 0$ at late times.
\end{proof}

There is an important example of the $f(R)$ theory that satisfies the conditions of Theorem 5, namely 
\begin{equation}
 f(R) = \alpha R^n
\end{equation}
defined on the interval $(0, \infty)$, where $\alpha$ is a positive constant and $n > 2$. Then $ f(R) - Rf'(R) = \alpha R^n (1 - n)$ is negative for all $R \in (0, \infty)$ since $n > 2$.  Now $$ \frac{f(R)-Rf'(R)}{(f'(R))^2} = \frac{1-n}{\alpha n^2}R^{2-n}$$ which, again since $n > 2$, tends to zero as $f'(R)$ tends to infinity.  So the condition (f1) is satisfied.  Condition (f2) is trivially satisified by noting that $2f(R) - Rf'(R) = \alpha R^n (2-n)$, which is negative for all $R \in (0,\infty)$.  The final condition that $$ \lim \sup \left(\frac{f(R)}{f(R) - Rf'(R)}\right) < \sqrt{3} - 1 $$ is shown to be true by noting that $$ \frac{f(R)}{f(R) - Rf'(R)} = \frac{1}{1-n} $$ which is, in fact, negative since $n > 2$.  Hence this particular form of $f(R)$ is an example of a function which satisfies the conditions of Theorem 5.  

This is an important example, since from the definitions of $V$ and $\phi$ in terms of $R$, we can show that this choice of $f(R)$ corresponds to the following form of potential: 
\begin{equation}
V(\phi) = -\frac{1}{16\pi}\left(\frac{\alpha (1-n)}{(\alpha n)^{n/n-1}}\right)e^{-8\pi \lambda \phi} 
\end{equation}
where $\lambda = \sqrt{\frac{2}{3}}\left(\frac{n-2}{n-1}\right)$, and since it can be shown that $\lambda < \sqrt{2}$ for all $n > 2$, we see that this is precisely the exponential potential described in \cite{slowroll} and in \cite{Halliwell}.

\section{$f(R)$ with matter}

Up to now, we have only considered cases where we have \emph{vacuum} solutions to the $f(R)$ field equations.  We would like to prove some results in the case where we add matter.  Suppose that $g$ is now a solution to the $f(R)$ field equations with matter described by the energy-momentum tensor $T^M_{\alpha\beta}$.  Then the action we consider is given by
\begin{equation}
  \mathcal{L} = \int  [f(R) + 8\pi L^M] \sqrt{-g} d^4x
\end{equation}
where $L^M$ is the Lagrangian density for the matter field.  Taking the variation of this action with respect to the metric gives the field equations for the \emph{$f(R)$-matter system}: 
\begin{equation} \label{fR+matter}
 f'(R)R_{\alpha\beta} - \frac{1}{2}g_{\alpha\beta}f(R) + \square (f'(R))g_{\alpha\beta} - \nabla_\alpha \nabla_\beta (f'(R)) = 8\pi T^M_{\alpha\beta}
\end{equation}
where the energy-momentum tensor, $T^M_{\alpha\beta}$ is defined by
\begin{equation}
 T^M_{\alpha\beta} = -\frac{\partial L^M}{\partial g^{\alpha\beta}} + \frac{1}{2}L^Mg_{\alpha\beta}
\end{equation}

The matter must also satisfy an equation of motion, which will depend on the type of matter used.  In general, the equation of motion will look like
\begin{equation}
 \frac{\partial L^M}{\partial \psi} - \nabla^\alpha \left(\frac{\partial L^M}{\partial (\nabla^\alpha \psi)}\right) = 0
\end{equation}
where $\psi$ is the matter field.
%
%
%
%
%

\subsection{$f(R)$ with matter and the coupled Einstein-scalar field-matter system}

In the case of $f(R)$ theory without matter, we found that the $f(R)$ theory has a one-to-one equivalence with the Einstein-scalar field system.  We can similarly show that when we add matter to the $f(R)$ theory, there is a correspondence between the $f(R)$-matter system and the \emph{coupled Einstein-scalar field-matter system}, which has been studied in \cite{Bieli}.  The paper \cite{Bieli} studied the coupled Einstein-scalar field-matter system and proved results on the late-time dynamics for certain choices of the potential of the scalar field.  By use of a conformal transformation we show here that there is a correspondence between the $f(R)$-matter system and the coupled Einstein-scalar field matter system and use the results in \cite{Bieli} to prove results on the late-time dynamics of solutions of the $f(R)$-matter system.

Let $\tilde{g}$ be a solution to the coupled Einstein-scalar field-matter equations.  Then $\tilde{g}$ must satisfy the following system of equations:
\begin{eqnarray} 
 \tilde{R}_{\alpha\beta} - \frac{1}{2}\tilde{R}\tilde{g}_{\alpha\beta} & = & 8\pi [\tilde{\nabla}_\alpha \phi \tilde{\nabla}_\beta \phi - \frac{1}{2}\tilde{\nabla}_\gamma \phi \tilde{\nabla}^\gamma \phi \tilde{g}_{\alpha\beta} - V(\phi) \tilde{g}_{\alpha\beta} \label{ESM1} \\ \nonumber
 & & + C(\phi)\tilde{T}^M_{\alpha\beta}] \\
\tilde{\square} \phi - V'(\phi) & = & -c(\phi) \tilde{T}^M \label{ESM2} \\
\tilde{\nabla}^\alpha \left( C(\phi)\tilde{T}^M_{\alpha\beta} \right) & = & c(\phi) \tilde{T}^M \tilde{\nabla}_\beta \phi \label{ESM3}
\end{eqnarray}
where $\tilde{R}_{\alpha\beta}$ and $\tilde{R}$ are the Ricci tensor and scalar curvature built from the metric $\tilde{g}$, $\tilde{\nabla}_\alpha$ is covariant differentiation with respect to $\tilde{g}$, $\phi$ is the scalar field with potential $V(\phi)$, $\tilde{T}^M_{\alpha\beta}$ is the energy-momentum tensor for the ordinary matter with $\tilde{T}^M$ its trace, $V'(\phi)$ is the derivative of $V$ with respect to $\phi$, $\tilde{\square}$ is the d'Alembertian operator and $C(\phi)$ and $c(\phi)$ are the coupling constants.

The energy-momentum tensor can be written as $$ \tilde{T}^M_{\alpha\beta}d\tilde{x}^\alpha d\tilde{x}^\beta = \tilde{S} - \tilde{j}\otimes d\tilde{t} - d\tilde{t} \otimes \tilde{j} + \tilde{\rho} d\tilde{t} \otimes d\tilde{t} $$ where $\tilde{S}_{\alpha\beta}$ is the spatial component of $\tilde{T}^M_{\alpha\beta}$, $\tilde{j}_\alpha$ is the current density and $\tilde{\rho}$ is the energy density.

We can now show the correspondence between the $f(R)$-matter system and the coupled Einstein-scalar field-matter system.  To do this, we let $g$ be a solution to the $f(R)$-matter field equations given in (\ref{fR+matter}).  The energy-momentum tensor $T^M_{\alpha\beta}$ is given (similarly as above) by 
\begin{equation} \label{emtensor}
 T^M_{\alpha\beta}dx^\alpha dx^\beta := S - j\otimes dt - dt\otimes j + \rho dt \otimes dt 
\end{equation}
We then make a conformal transformation as follows (c.f. section 5 in \cite{Bieli}).  
\begin{equation*}
 g \mapsto \tilde{g} := Fg \hspace{.5in} t \mapsto \tilde{t} := \int_{t_0}^t F^{1/2} dt 
\end{equation*}
\begin{equation*}
 S \mapsto \tilde{S} := S \hspace{.5in} j \mapsto \tilde{j} := F^{-1/2}j \hspace{.5in} \rho \mapsto \tilde{\rho} := F^{-1} \rho
\end{equation*}
Then we see that the components of the energy-momentum tensor, $\tilde{T}^M$, in the $\tilde{g}$ coordinate system are the same as the components of the energy-momentum tensor, $T^M$, in the $g$ coordinate system.

Using  this transformation and equation (\ref{fR+matter}), we find for the $\tilde{g}$ system that the following holds (c.f. equation (\ref{tran}) for the vacuum case)
\begin{eqnarray*}
 \tilde{R}_{\alpha\beta} - \frac{1}{2}\tilde{R}\tilde{g}_{\alpha\beta} & = & 8\pi F^{-1}\tilde{T}^M_{\alpha\beta} + \frac{1}{2}f(R)F^{-2}\tilde{g}_{\alpha\beta} - \frac{1}{2}F^{-1}R\tilde{g}_{\alpha\beta} \nonumber \\
 & & + \frac{3}{2}F^{-2}\tilde{\nabla}_\alpha F\tilde{\nabla}_\beta F - \frac{3}{4}F^{-2}\tilde{\nabla}^\gamma F\tilde{\nabla}_\gamma F\tilde{g}_{\alpha\beta}
\end{eqnarray*}
where $\tilde{R}_{\alpha\beta}$ and $\tilde{R}$ are the Ricci tensor and scalar curvature built from the metric $\tilde{g}$.  If we then make the choices (as in the vacuum case)
\begin{equation} \label{phimatter}
 \phi = \sqrt{\frac{3}{16\pi}} \ln F
\end{equation}
and
\begin{equation} \label{Vmatter}
 V(\phi) = -\frac{1}{16\pi}\frac{f(R) - RF}{F^2}
\end{equation}
as well as the choice
\begin{equation} \label{C}
 C(\phi) = F^{-1}
\end{equation}
we find that this equation becomes
\begin{equation}
 \tilde{R}_{\alpha\beta} - \frac{1}{2}\tilde{R}\tilde{g}_{\alpha\beta} = 8\pi \left( \tilde{\nabla}_\alpha \phi \tilde{\nabla}_\beta \phi - \frac{1}{2}\tilde{\nabla}_\gamma \phi \tilde{\nabla}^\gamma \phi \tilde{g}_{\alpha\beta} - V(\phi ) \tilde{g}_{\alpha\beta} + C(\phi) \tilde{T}^M_{\alpha\beta}\right)
\end{equation}
This is the first equation for the coupled Einstein-scalar field-matter system.  

Now consider equation (\ref{fR+matter}) and take its trace.  We get
\begin{equation*}
 f'(R)R - 2f(R) + 3\square F = 8\pi T^M
\end{equation*}
Multiply through by $\frac{1}{16\pi}\sqrt{\frac{16\pi}{3}}F^{-2}$ to get
\begin{equation} \label{traceM}
 \frac{1}{16\pi}\sqrt{\frac{16\pi}{3}}\left[\frac{f'(R)R - 2f(R)}{F^2}\right] + \frac{3}{16\pi}\sqrt{\frac{16\pi}{3}}F^{-2}\square F = \frac{1}{2}\sqrt{\frac{16\pi}{3}}F^{-2}T^M
\end{equation}
If we make the conformal transformation $g \mapsto \tilde{g}$ and $T^M \mapsto \tilde{T}^M$ as before, we find that the wave operator on $F$ transforms like 
\begin{equation*}
 \square F = F \tilde{\square}F - \tilde{\nabla}^\alpha F \tilde{\nabla}_\alpha F
\end{equation*}
Recalling that $F = e^{\sqrt{\frac{16\pi}{3}}\phi}$ then gives
\begin{eqnarray*}
 \square F & = & \frac{16\pi}{3}F^2 \tilde{\nabla}^\gamma \phi \tilde{\nabla}_\gamma \phi + \sqrt{\frac{16\pi}{3}}F^2 \tilde{\square} \phi - \frac{16\pi}{3}F^2 \tilde{\nabla}^\alpha \phi \tilde{\nabla}_\alpha \phi \nonumber \\
 & = & \sqrt{\frac{16\pi}{3}}F^2 \tilde{\square} \phi
\end{eqnarray*}
Then substituting into (\ref{traceM}), writing $T^M$ as $F \tilde{T^M}$ (since $T^M \mapsto \tilde{T}^M = \tilde{T}^M_{\alpha\beta} \tilde{g}^{\alpha\beta} = F^{-1}T^M_{\alpha\beta}g^{\alpha\beta} = F^{-1}T^M$), and recalling the definition of the derivative of the potential, $V(\phi)$, of the scalar field  given in (\ref{potderiv}) gives
\begin{equation*}
 -V'(\phi) + \tilde{\square} \phi = \frac{1}{2}\sqrt{\frac{16\pi}{3}}F^{-1}\tilde{T}^M
\end{equation*}
which gives equation (\ref{ESM2}), where the coupling constant $c(\phi)$ is given by 
\begin{equation} \label{c}
 c(\phi) = -\frac{1}{2}\sqrt{\frac{16\pi}{3}}F^{-1}
\end{equation}

To recover the third (and final) equation, given by (\ref{ESM3}), we must use the fact that the energy-momentum tensor of the matter in the $f(R)$ theory is divergence-free i.e.
\begin{equation}
 \nabla^\alpha T^M_{\alpha\beta} = 0
\end{equation}
The definition of the covariant derivative of a tensor gives
\begin{equation*}
 \nabla_\gamma T^M_{\alpha\beta} = \partial_\gamma T^M_{\alpha\beta} - \Gamma^\delta_{\gamma\alpha} T^M_{\delta\beta} - \Gamma^\delta_{\gamma\beta} T^M_{\alpha\delta}
\end{equation*}
Again, we can make the conformal transformation $g \mapsto \tilde{g}, T^M \mapsto \tilde{T}^M$.  We get the following
\begin{eqnarray*}
 \nabla^\alpha T^M_{\alpha\beta} & = & F\tilde{g}^{\alpha\gamma} \left[ \tilde{\partial}_\gamma \tilde{T}^M_{\alpha\beta} - \left( \tilde{\Gamma}^\delta_{\gamma\alpha} - \frac{1}{2}F^{-1}(\tilde{\partial}_\alpha F \tilde{\delta}^\delta_\gamma + \tilde{\partial}_\gamma F\tilde{\partial}^\delta_\alpha - \tilde{\partial}^\delta F\tilde{g}_{\gamma\alpha})\right) \tilde{T}^M_{\delta\beta} \right] \nonumber \\
 & & - F\tilde{g}^{\alpha\gamma} \left[ \left( \tilde{\Gamma}^\delta_{\gamma\beta} - \frac{1}{2}F^{-1}(\tilde{\partial}_\beta F \tilde{\delta}^\delta_\gamma + \tilde{\partial}_\gamma F\tilde{\partial}^\delta_\beta - \tilde{\partial}^\delta F\tilde{g}_{\gamma\beta})\right) \tilde{T}^M_{\alpha\delta} \right]
\end{eqnarray*}
which simplifies to
\begin{equation*}
 \nabla^\alpha T^M_{\alpha\beta} = F \tilde{\nabla}^\alpha \tilde{T}^M_{\alpha\beta} - \tilde{\partial}^\gamma F \tilde{T}^M_{\gamma\beta} + \frac{1}{2}\tilde{\partial}_\beta F \tilde{T}^M
\end{equation*}
and since we require that the divergence of $T^M_{\alpha\beta}$ is zero, this gives
\begin{equation*}
 F \tilde{\nabla}^\alpha \tilde{T}^M_{\alpha\beta} - \tilde{\nabla}^\gamma F \tilde{T}^M_{\gamma\beta} + \frac{1}{2} \tilde{\nabla}_\beta F\tilde{T}^M = 0
\end{equation*}
Since we know already that $C(\phi) = F^{-1}$, we see that 
\begin{equation*}
 F^2 \tilde{\nabla}^\alpha \left(C(\phi)\tilde{T}^M_{\alpha\beta}\right) = F \tilde{\nabla}^\alpha \tilde{T}^M_{\alpha\beta} - \tilde{\nabla}^\alpha F \tilde{T}^M_{\alpha\beta}
\end{equation*}
and so we find that
\begin{equation*}
 F^2 \tilde{\nabla}^\alpha \left( C(\phi) \tilde{T}^M_{\alpha\beta}\right) = -\frac{1}{2}\tilde{\nabla}_\beta F\tilde{T}^M
\end{equation*}
Recalling that $F = e^{\sqrt{\frac{16\pi}{3}}\phi}$ then gives
\begin{equation*}
 \tilde{\nabla}^\alpha \left(C(\phi)\tilde{T}^M_{\alpha\beta}\right) = -\frac{1}{2}\sqrt{\frac{16\pi}{3}}F^{-1}\tilde{\nabla}_\beta \phi \tilde{T}^M
\end{equation*}
which, recalling the definition of $c(\phi)$,  gives the third and final condition for $\tilde{g}$ to be a solution to the Einstein-scalar field-matter system.

Therefore, if the metric $g_{\alpha\beta}$ is a solution to the $f(R)$-matter field equations, we find that under suitable choices of $\phi$ and $V(\phi)$ and the coupling constants $C(\phi)$ and $c(\phi)$, given in equations (\ref{phimatter}), (\ref{Vmatter}), (\ref{C}) and (\ref{c}) the metric $\tilde{g}_{\alpha\beta} = Fg_{\alpha\beta}$ is a solution to the coupled Einstein-scalar field-matter system described by equations (\ref{ESM1}), (\ref{ESM2}) and (\ref{ESM3}).  We can thus apply the results given in \cite{Bieli} on the dynamics of solutions to the coupled Einstein-scalar field-matter system and use them to prove results on the dynamics of the $f(R)$-matter system.

\subsection{The potential admits a strictly positive minimum}

As in the vacuum case, we will consider first the situation where the potential admits a strictly positive minimum.  We can use the results of \cite{Bieli} to prove results for the $f(R)$-matter system.

We would first like to show, as in the vacuum case, that we can change the assumptions used in \cite{Bieli} on the potential of the scalar field to prove an analogous theorem for the Einstein-scalar field-matter system.  We show that the following Theorem holds (analogous to Proposition 3 in \cite{Bieli}):

\begin{theorem}
 Consider a solution, $\tilde{g}$, of the coupled Einstein-scalar field-matter system, given by equations (\ref{ESM1}), (\ref{ESM2}), (\ref{ESM3}), of Bianchi type I-VIII.  Suppose that the energy-momentum tensor, $\tilde{T}_{\alpha\beta}$, satisfies the assumptions (as in \cite{Bieli})

(DEC) $\tilde{j}_a \tilde{j}_b \tilde{g}^{ab} \leq \tilde{\rho}^2, \hspace{.1in} \tilde{S}_{ab}\tilde{g}^{ab} \leq 3\tilde{\rho}$ 

(SEC) $\tilde{\rho} + \tilde{S}_{ab}\tilde{g}^{ab} \geq 0$ \newline
the coupling constants satisfy (as in \cite{Bieli})

(C) $|c(\phi)| \leq C_0C(\phi)$ \newline
and the potential, $V$, of the scalar field is of class $C^2$ and satisfies (alternative assumptions to \cite{Bieli})

(V1) $V(\phi) \geq V_0$ for a constant $V_0 > 0$

(V2) $V$ tends to $\infty$ as $\phi$ tends to its endpoints. \newline
If the solution is initially expanding ($\tilde{H} > 0$) then, as $t \rightarrow \infty$, the following limits are attained:

(i) $\tilde{H} \rightarrow \tilde{H}_\infty$, with $\tilde{H}_\infty \geq \tilde{H}_0 > 0$

(ii) $V \rightarrow V_\infty$, where $V_\infty = \frac{3}{8\pi}\tilde{H}_\infty^2$

(iii) $V' \rightarrow 0$

(iv) $\dot{\phi}, \ddot{\phi} \rightarrow 0$ \newline
where $\tilde{H}$ is the Hubble parameter for the coupled Einstein-scalar field-matter system.
\end{theorem}

\begin{proof}
 First note that the proof of Proposition 3 in \cite{Bieli} first requires proving Lemma 1 and Proposition 2 in \cite{Bieli}.  To prove these for our alternative assumptions on the potential of the scalar field, we must only find where assumptions (P1), (P2) and (P3) were used and check that we get the same results with assumptions (V1) and (V2).  For Lemma 1, it was only necessary to use (P1) to show that $V$ is positive.  This is clearly obtained from our alternative assumption (V1).  (P2) was used to give the bound on $V'(\phi)$.  From our assumption (V2), however, as in the case for vacuum $f(R)$, we see that, since $V$ is bounded to the future, we must have that $\phi$ remains bounded away from the endpoints of the domain of $V$ as $t \rightarrow \infty$.  Then, since $V$ is smooth, we find that $V'$ must also be bounded to the future.  The rest of the proof of Lemma 1 goes through as in \cite{Bieli}.  Proposition 2 does not require any of the assumptions (P1), (P2), (P3) and so the proof of this is identical to that in \cite{Bieli}.  With the results of Lemma 1 and Proposition 2 at our disposal, it only remains to show how Proposition 3 can be proved with our alternative assumptions on the potential.  

We first note that parts (i) and (ii) are proved in the same way as in \cite{Bieli}, since they do not require (P1), (P2) or (P3).  For part (iii), \cite{Bieli} requires (P3) to show that, in the case where $\phi$ converges as $t \rightarrow \infty$, if $\phi$ converges to plus or minus infinity, then we still have that $V'$ converges as $t \rightarrow \infty$.  With our assumptions (V1) and (V2), however, we know already that $\phi$ remains bounded as $t \rightarrow \infty$, which rules out the possibility that $\phi$ converges to $\pm \infty$, and so we no longer require this assumption.  The rest of the proof then follows as in \cite{Bieli}.  It should be noted that the discrepancy in the constants in part (ii) arise from the fact that $8\pi G = 1$ in \cite{Bieli}.
\end{proof}

We are now ready to transfer these results onto the $f(R)$-matter system.  Recall the discussion from the vacuum case in section 3.2, which gave the conditions on the $f(R)$ theory that are equivalent to the condition that the potential of the scalar field blows up at its endpoints.  These conditions are labelled (F1), (F2) and (F3).  With this in mind, we can now prove the following theorem on the dynamics of the $f(R)$-matter system.

\begin{theorem}
 Consider a solution, $g_{\alpha\beta}$, of the $f(R)$-matter system of Bianchi type I-VIII.  Suppose that the energy-momentum tensor, $T^M_{\alpha\beta}$, given by (\ref{emtensor}), satisfies

(1) $j_aj_bg^{ab} \leq \rho^2$, \hspace{.1in} $S_{ab}g^{ab} \leq 3\rho$

(2) $\rho + S_{ab}g^{ab} \geq 0$ \newline
and that the function $f(R)$ satisfies 

(3) $f(R) - Rf'(R) < 0$ for all $f'(R) > 0$.

(4) $f(R) - Rf'(R)$ satisfies the conditions (F1), (F2) and (F3) given in section 3.2. \newline
If the solution satisfies $H + \frac{\dot{F}}{F} > 0$ initially and exists globally to the future then, as $t \rightarrow \infty$, the following limits are attained:

(i) $H \rightarrow H_\infty$ with $H_\infty \geq \tilde{H}_0 > 0$

\vspace{.1in}
(ii) $-\frac{f(R) - Rf'(R)}{(f'(R))^2} \rightarrow K$ where $K = 6H_\infty^2$

\vspace{.1in}
(iii) $\frac{2f(R) - Rf'(R)}{(f'(R))^2} \rightarrow 0$ \newline
where $H$ is the Hubble parameter for the $f(R)$-matter system.
\end{theorem}

\begin{proof}
 From the relationships between $f(R)$ and the potential, $V(\phi)$, of $\phi$, using the discussion preceding Theorem 2 in Section 3.2, and from the relationship (\ref{Htilde}) between the Hubble parameters, we see that, if $g$ is a solution to the $f(R)$-matter system satisfying assumptions (3) and (4) of Theorem 7, then $\tilde{g} = Fg$ is a solution to the coupled Einstein-scalar field-matter system, where the potential, $V$, satisfies (V1), (V2) of Theorem 6.  In addition, if the matter of the $f(R)$-matter system satisfies (1) and (2), then we see from the conformal transformation described in the previous section that the matter, $\tilde{T}$, of the Einstein-scalar field-matter system satisfies (DEC), (SEC).  Lastly, from the definitions of the coupling constants $C(\phi)$ and $c(\phi)$ (given in (\ref{C}), (\ref{c})), we see that the ratio between them is constant.  Hence the condition (C) of Theorem 6 is automatically satisfied.

Therefore, if $g$ is a solution to the $f(R)$-matter system satisfying (1) - (4) of Theorem 7, then $\tilde{g} = Fg$ is a solution to the coupled Einstein-scalar field-matter system satisfying (DEC), (SEC), (V1), (V2) and (C) in Theorem 6.  So the results of Theorem 6 hold for $\tilde{g}$.  Then parts (ii) and (iii) of Theorem 7 are proved directly from the definitions of $V(\phi)$ and $V'(\phi)$ in terms of $R, f(R)$ and $f'(R)$, although we must still prove the relationship between $K$ and $H_\infty$.  

For part (iii), recall from (\ref{Htilde}) that $\tilde{H} = H + \frac{\dot{F}}{F}$ and that $\dot{\phi} = \sqrt{\frac{3}{16\pi}}\frac{\dot{F}}{F}$.  From Theorem 6, we know that $\dot{\phi} \rightarrow 0$ as $t \rightarrow \infty$.  Hence we find that $H$ behaves like $\tilde{H}$ at late times and so $H \rightarrow H_\infty (= \tilde{H}_\infty)$ with $H_\infty \geq \tilde{H}_0 > 0$.  

To show the relationship between $K$ and $H_\infty$, note that the quantity $$-\frac{f(R) - Rf'(R)}{(f'(R))^2}$$ is equal to $16\pi V$ and so tends to $16\pi V_\infty$ which, from Theorem 6, is equal to $6\tilde{H}_\infty^2$.  Since we have shown above that $\tilde{H}_\infty = H_\infty$, this completes the proof.
\end{proof}

\subsection{Intermediate Inflation}

As in the vacuum case, it has also been proven for the coupled Einstein-scalar field-matter system \cite{Bieli} that accelerated expansion occurs for the ``intermediate inflation'' case, where the potential of the scalar field is strictly positive, but tends to zero as $\phi$ tends to infinity.  Once again, we can use results from \cite{Bieli} on the dynamics of the Einstein-scalar field-matter system to prove results on the dynamics of the $f(R)$-matter system.  We assume, as in section 3.3 for the vacuum intermediate inflation case, that $f'(R) \rightarrow \infty$ as $t \rightarrow \infty$. We prove the following theorem:

\begin{theorem}
 Let $g$ be a solution of the $f(R)$-matter system of Bianchi type I-VIII.  Suppose that the energy-momentum tensor, $T^M_{\alpha\beta}$, satisfies

(1) $j_aj_bg^{ab} \leq \rho^2, S_{ab}g^{ab} \leq 3\rho$

(2) $\rho + S_{ab}g^{ab} \geq 0$ \newline
and that the function $f(R)$ satisfies

(3) $ f(R) - Rf'(R) < 0$ with $\frac{f(R) - Rf'(R)}{(f'(R))^2} \rightarrow 0$ as $f'(R) \rightarrow \infty$

\vspace{.1in}
(4) $ 2f(R) - Rf'(R) < 0$ for all $R$

\vspace{.1in}
(5) $ -1 < \lim \sup \left(\frac{f(R)}{f(R) - Rf'(R)}\right) < 0$ as $f'(R) \rightarrow \infty$ \vspace{.1in} \newline
If the solution satisfies $H + \frac{\dot{F}}{F} > 0$ initially and exists globally to the future then the solution undergoes accelerated expansion at late times.
\end{theorem}

\begin{proof}
 Once again we use the relationship between the $f(R)$-matter system and the coupled Einstein-scalar field-matter system and the discussion in section 3.3 of this paper to show that, if $g$ is a solution to the $f(R)$-matter system satisfying the assumptions of the theorem, then $\tilde{g} = Fg$ is a solution to the coupled Einstein-scalar field-matter system satisfying the conditions of section 4 of \cite{Bieli}.  In particular, for the $\tilde{g}$ solution, the result of Proposition 5 in \cite{Bieli} proves that, if $\alpha := \lim \sup (-V'/V) < \sqrt{\frac{16\pi}{3}}$, then the following bound is obtained: $$ \lim \sup \left(\frac{3\tilde{H}^2}{8\pi V}\right) \leq \left( 1 - \frac{1}{2}\sqrt{\frac{1}{12\pi}}\alpha\right)^{-1} $$  The requirement of the bound on the ratio $V'/V$ is deduced from assumption (5) of Theorem 8 above.  This then gives $$ \lim \sup \left(\frac{3\tilde{H}^2}{8\pi V}\right) < 3/2 $$ for sufficiently large $t$ which, as we saw in section 3.3, is equivalent to $$ \dot{\tilde{H}} + \tilde{H}^2 \geq 0 $$ which gives $\ddot{\tilde{a}} > 0$ and therefore accelerated expansion at late times.  It should be noted once more that there is a slight discrepancy between the factors used in this paper and those used in \cite{Bieli}, due to the notation $8\pi G = 1$ in \cite{Bieli}.

Similarly to section 3.3, we must use the relationship between the Hubble parameters of the two systems and information gained during section 4 of \cite{Bieli}, to show that the expression $\dot{H} + H^2$ behaves like $\dot{\tilde{H}} + \tilde{H}^2$ as $t \rightarrow \infty$, which proves accelerated expansion for the $f(R)$-matter system.  This is done by noting that the discussion after Proposition 4 in \cite{Bieli} tells us that $\tilde{H} \rightarrow 0$ as $t \rightarrow \infty$.  Then this, together with the decay results of Proposition 4 and the equation (13) of \cite{Bieli} gives a bound on $\frac{\dot{\phi}}{\tilde{H}}$, which tells us that $\dot{\phi} \rightarrow \infty$ as $t \rightarrow \infty$.  The bound on $V'/V$ and the assumption that $V \rightarrow 0$ as $t \rightarrow \infty$ tells us that $V' \rightarrow 0$ as $t \rightarrow \infty$.  Then the equation of motion $$ \ddot{\phi} + 3\tilde{H}\dot{\phi} + V'(\phi) = c(\phi)\tilde{T}$$ and the information above gives $\ddot{\phi} \rightarrow 0$ as $t \rightarrow \infty$.  Then (c.f. section 3.3), we see that $H$ and $\dot{H}$ behave like $\tilde{H}$ and $\dot{\tilde{H}}$ as $t \rightarrow \infty$, which gives the result that $\dot{H} + H^2 > 0$ for sufficiently large $t$, which concludes the proof that the solution undergoes accelerated expansion at late times.
\end{proof}

\section{Conclusions}

In conclusion, we have shown that it is possible to exploit the connection between the $f(R)$ theory and the Einstein-scalar field system to prove mathematically rigorous results on the late-time dynamics of the $f(R)$ theory.  In particular, we have shown that, for a Bianchi type I-VIII space-time, by placing certain restrictions on the function $f(R)$ of the $f(R)$ theory (obtained according to restrictions placed on the potential, $V$, of the scalar field in the Einstein-scalar field system), we have proved that solutions to the $f(R)$ theory satisfying these restrictions undergo accelerated expansion at late times.  We have shown that this is true in both the vacuum case and the case where we add ordinary matter to the $f(R)$ theory. 

\section*{Acknowledgements}

The author would like to thank Alan D. Rendall for many valuable discussions.

\bibliography{fRpaper}

\bibliographystyle{unsrt}

\end{document}